\def\aa{{A\&A}}

\def\aj{{AJ}}
\def\annrev{{ARA\&A}}
\def\apj{{ApJ}}
\def\apjs{{ApJS}}

\def\mnras{{MNRAS}}
\def\nat{{Nature}}
\def\pasp{{PASP}}

\newcommand{\hst}{\emph{HST}}
\newcommand{\etal}{{et al.}}
\newcommand{\mbh}{\ensuremath{M_\bullet}}
\newcommand{\sigmastar}{\ensuremath{\sigma_\star}}
\newcommand{\msigma}{\ensuremath{\mbh-\sigmastar}}
\newcommand{\kms}{km s\ensuremath{^{-1}}}
\newcommand{\rg}{\ensuremath{r_\mathrm{G}}}
\newcommand{\msun}{\ensuremath{M_\odot}}
\newcommand{\hal}{H$\alpha$}
\newcommand{\hbeta}{H$\beta$}
\newcommand{\arcdeg}{\mbox{$^\circ$}}% 
\newcommand{\per}{\ensuremath{^{-1}}}
\newcommand{\water}{H$_2$O}
\newcommand{\sgas}{\ensuremath{\sigma_{\mathrm{gas}}}}
\newcommand{\lbul}{\ensuremath{L_\mathrm{bulge}}}
\newcommand{\mbul}{\ensuremath{M_\mathrm{bulge}}}
\newcommand{\rblr}{\ensuremath{r_\mathrm{BLR}}}

\newcommand{\tlag}{\ensuremath{t_\mathrm{lag}}}
\def\arcsec{\hbox{\ensuremath{^{\prime\prime}}}}
\def\farcs{\hbox{\ensuremath{.\!\!^{\prime\prime}}}}
\newcounter{ctr}
\def\ion#1#2{\setcounter{ctr}{#2}#1$\;${\Roman{ctr}}\relax}

\documentclass[cup5b]{caps}
\usepackage{graphicx}
\usepackage{amssymb}
\usepackage{ociwsymp1}
\HeadText{A. J. Barth}

\def\plotone#1{\centering \leavevmode
\includegraphics[width=.95\columnwidth]{#1}}

\begin{document}

\pagenumbering{arabic}

\author[]{AARON J. BARTH\\ California Institute of Technology}

\chapter{Black Holes in Active Galaxies}

\begin{abstract}

Recent years have seen tremendous progress in the quest to detect
supermassive black holes in the centers of nearby galaxies, and
gas-dynamical measurements of the central masses of active galaxies
have been valuable contributions to the local black hole census.  This
review summarizes measurement techniques and results from observations
of spatially resolved gas disks in active galaxies, and reverberation
mapping of the broad-line regions of Seyfert galaxies and quasars.
Future prospects for the study of black hole masses in active
galaxies, both locally and at high redshift, are discussed.

\end{abstract}

\section{Introduction}

The detection of supermassive black holes in the nuclei of many nearby
galaxies has been one of the most exciting discoveries in
extragalactic astronomy during the past decade.  Accretion onto black
holes has long been understood as the best explanation for the
enormous luminosities of quasars (Salpeter 1964; Zel'dovich \& Novikov 
1964; Rees 1984), and the luminosity generated by quasars over the
history of the Universe implies that most large galaxies must contain
a black hole as a relic of an earlier quasar phase (So\l tan 1982;
Chokshi \& Turner 1992; Small \& Blandford 1992).  While the search
for evidence of black holes in nearby galaxies began 25 years ago with
the seminal studies of M87 by Sargent \etal\ (1978) and Young \etal\
(1978), only a handful of galaxies were accessible to such
measurements until the repair of the \emph{Hubble Space Telescope}
(\hst) in 1993 made it possible to study the central dynamics of
galaxies routinely at 0\farcs1 resolution.  In addition to the recent
dynamical searches for black holes in active and inactive galaxies
with \hst, the existence of black holes has been further confirmed by
ground-based observations of the Galactic Center (see Ghez, this
volume) and by radio observations of the H$_2$O maser disk in the
Seyfert 2 galaxy NGC 4258 (Miyoshi \etal\ 1995).  As the evidence for
supermassive black holes in galaxy centers has strengthened, it has
become clear that nuclear activity and the growth of black holes must
be integral components of the galaxy formation process.

\begin{figure}
\centering
%\scalebox{0.45}{\includegraphics{msigma.ps}}
\plotone{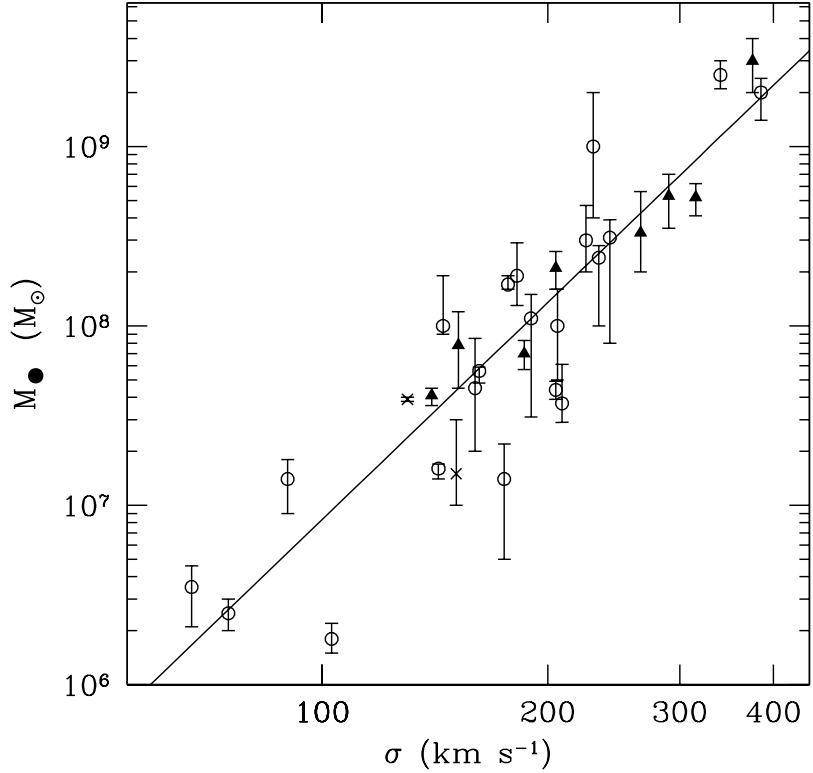}
\caption{The correlation between black hole mass and stellar velocity
  dispersion.  Triangles denote galaxies measured with \hst\
  observations of gas dynamics, crosses are H$_2$O maser galaxies, and
  circles denote stellar-dynamical detections.  The diagonal line is
  the best fit to the data as determined by Tremaine \etal\ (2002).}
\label{msigma}
\end{figure}

With measurements of black hole masses in several galaxies, it became
possible for the first time to study the demographics of the black
hole population and the connection of the black holes with their host
galaxies.  Kormendy \& Richstone (1995) showed that \mbh\ was
correlated with \lbul, the luminosity of the spheroidal ``bulge''
component of the host galaxy, albeit with substantial scatter.  More
intriguing was the discovery that \mbh\ is very tightly correlated
with \sigmastar, the stellar velocity dispersion in the host galaxy
(Ferrarese \& Merritt 2000; Gebhardt \etal\ 2000a).  The scatter in
this relation is surprisingly small; Tremaine \etal\ (2002) estimate
the dispersion to be $<0.3$ dex in log \mbh\ at a given value of
\sigmastar.  This is a remarkable finding: it implies that the masses
of black holes, objects that inhabit scales of $\lesssim10^{-4}$ pc in
galaxy nuclei, are almost \emph{completely} determined by the bulk
properties of their host galaxies on scales of hundreds or thousands
of parsecs.  Although the \msigma\ correlation is well established,
its slope, and the amount of intrinsic scatter, remain somewhat
controversial.  The currently available sample of galaxies with
accurate determinations of \mbh\ is still modest.  More measurements
of black hole masses in nearby galaxies are needed, over the widest
possible range of host galaxy types and velocity dispersions, in order
to obtain a definitive present-day black hole census.

Gas-dynamical measurements of black hole masses in active galactic
nuclei (AGNs) are an essential contribution to this pursuit, as
illustrated in Figure \ref{msigma}. \hst\ observations of ionized gas
disks are vitally important for tracing the upper end of the black
hole mass function, where stellar-dynamical measurements are hampered
both by the low stellar surface brightness of the most massive
elliptical galaxies and by the possibility of velocity anisotropy in
nonrotating ellipticals.  Observations of maser emission from
molecular disks in active galaxies have provided the most solid black
hole detection outside of our own Galaxy, strengthening the case that
the massive dark objects discovered in \hst\ surveys are indeed likely
to be supermassive black holes.  Reverberation mapping, and secondary
methods that are calibrated by comparison with the reverberation
technique, offer the most promising methods to determine black hole
masses at high redshift.

The topic of black holes in active galaxies is vast, and this review
will only concentrate on gas-dynamical measurements of black hole
masses in AGNs.  Before discussing the methods and results, a few
general comments are in order.  As Kormendy \& Richstone (1995) have
pointed out, there is a potentially serious drawback to any
measurement technique based on gas dynamics: unlike stars, gas can
respond to nongravitational forces, and the motions of gas clouds do
not always reflect the underlying gravitational potential.  For all
methods based on gas dynamics, it is absolutely crucial to verify that
the gas is actually in gravitational orbits about the central mass.
If, for example, AGN-driven outflows or other nongravitational
motions dominate, then black hole masses derived under the assumption
of gravitational dynamics will be seriously compromised or completely
erroneous.  With that said, there are now numerous examples of ionized
gas disks, and at least one maser disk, that clearly show orderly
circular rotation.  For reverberation mapping, the dynamical state of
the broad-line emitting gas is more difficult to ascertain, but as
discussed in \S\ref{reverb} below, recent observations have provided
some encouragement.

It must also be emphasized that, while these measurement techniques
are capable of detecting dark mass concentrations in the centers of
galaxies and determining their masses with varying degrees of
accuracy, the observations do not actually prove that the dark mass is
in the form of a supermassive black hole.  The spatial resolution of
gas-dynamical observations with \hst\ typically corresponds to
$\sim10^{5-6}$ Schwarzschild radii.  This is often sufficient to
resolve the region over which the black hole dominates the
gravitational potential of its host galaxy, but optical techniques are
incapable of resolving the region in which relativistic motion occurs
in the strong gravitational field near the black hole's event horizon.
The conclusion that the massive dark objects detected in nearby
galaxies are actually black holes is supported by the two most
convincing dynamical detections, in our own Galaxy and in NGC 4258; in
both objects the density of the central dark mass is inferred to be so
large that reasonable alternatives to a black hole can be ruled out
(Maoz 1998; Ghez, this volume).  The best evidence for highly relativistic 
motion in the inner accretion disks of AGNs comes from X-ray spectra showing
extremely broadened ($\sim0.3c$), gravitationally redshifted Fe K line
emission in Seyfert nuclei (Tanaka \etal\ 1995; Nandra \etal\ 1997).
While this signature has only been convincingly detected in a handful
of objects, it offers a powerful confirmation of the AGN paradigm, and
analysis of the relativistically broadened line profiles may even
reveal evidence for the black hole's spin (e.g., Iwasawa \etal\ 1996).

\section{Black Hole Masses from Dynamics of Ionized Gas Disks}

A striking discovery from the first years of \hst\ observations was
the presence of round, flattened disks of ionized gas and dust in the
centers of some nearby radio galaxies (Jaffe \etal\ 1993; Ford \etal\
1994).  It was previously known from ground-based imaging that many
ellipticals contained nuclear patches of dust (Kotanyi \& Ekers 1979;
Sadler \& Gerhard 1985; Ebneter, Davis, \& Djorgovski 1988), but \hst\
revealed that the dust was often arranged in well-defined disks too
small to be resolved from the ground.  Imaging surveys with \hst\ have
found such disks, with typical radii of $100-1000$ pc, in $\sim20\%$
of giant elliptical galaxies (e.g., van~Dokkum \& Franx 1995; Verdoes
Kleijn \etal\ 1999; Capetti \etal\ 2000; de~Koff \etal\ 2000; Tomita
\etal\ 2000; Tran \etal\ 2001; Laine \etal\ 2003).

\begin{figure}
\centering
\scalebox{0.9}{\includegraphics{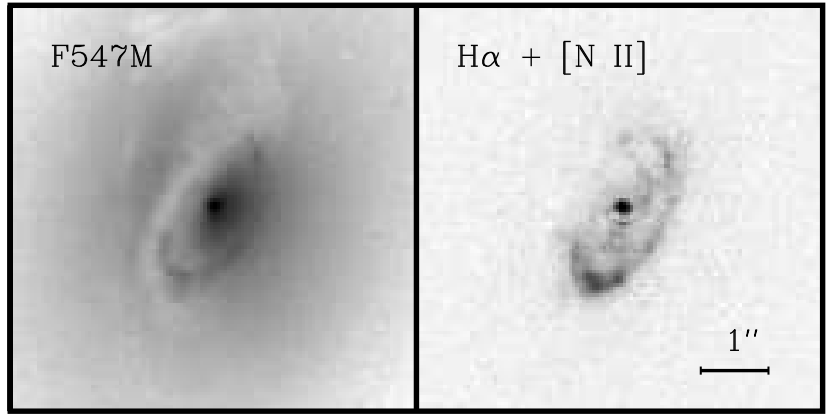}}
\caption{An example of a dusty emission-line disk: \hst\ images of the
   S0 galaxy NGC 3245.  The left panel is a continuum image taken with
   the \hst\ F547M filter (equivalent to the $V$ band), and the right
   panel shows a continuum-subtracted, narrow-band image isolating the
   \hal\ and [\ion{N}{2}] emission lines.  At $D=21$ Mpc, 1\arcsec\
   corresponds to 100 pc.}
\label{ngc3245image}
\end{figure}

Jaffe \etal\ (1999) show that nuclear gas disks in early-type galaxies
fall into two general categories.  The most common type are dusty
disks, which are easily detected by their obscuration in broad-band
optical \hst\ images.  The dust is usually accompanied by an ionized
component.  Figure \ref{ngc3245image} shows an example, the disk in
the S0 galaxy NGC 3245.  The second class consists of ionized gas
without associated dust disks.  M87 (Ford \etal\ 1994) is the
prototype of this category.  Ionized disks are sometimes found to have
filamentary or spiral structure, and patches of dust may be present as
well.  A comprehensive study of disk orientations in radio galaxies by
Schmitt \etal\ (2002) finds that the radio jets are not preferentially
aligned along the disk rotation axis, although jets tend not to be
oriented close to the disk plane.

Soon after the first \hst\ servicing mission, the first spectroscopic
investigations of the kinematics of these disks were performed with
the Faint Object Spectrograph (FOS).  The first target was M87, for
which Harms \etal\ (1994) detected a steep velocity gradient across
the nucleus in the \hal, [\ion{N}{2}], and [\ion{O}{3}] emission
lines, consistent with Keplerian rotation in an inclined disk.  The
central mass was found to be $(2.4 \pm 0.7) \times 10^9$ \msun,
remarkably close to the values first determined by Young \etal\ (1978)
and Sargent \etal\ (1978).  The second gas-dynamical study with \hst\
found a central dark mass of $(4.9\pm1.0)\times10^8$ \msun\ in the
radio galaxy NGC 4261 (Ferrarese, Ford, \& Jaffe 1996).  These
dramatic results opened a new chapter in the search for supermassive
black holes, demonstrating that spatially resolved gas disks could
indeed be used to measure the central masses of galaxies.  In contrast
to stellar dynamics, the gas-dynamical method is extremely appealing
in its simplicity.  Modeling the kinematics of a thin, rotating disk
is conceptually straightforward.  Furthermore, observations of
emission-line velocity fields require less telescope time than
absorption-line spectroscopy.  Since the FOS was a single-aperture
spectrograph, however, it was not well-suited to the task of mapping
out emission-line velocity fields in detail, and FOS gas-dynamics data
were only obtained for a few additional galaxies (van~der~Marel \& 
van~den~Bosch 1998; Ferrarese \& Ford 1999; Verdoes Kleijn \etal\ 2000).

After the initial FOS detections, progress was made on two fronts.
The installation of the Space Telescope Imaging Spectrograph (STIS), a
long-slit instrument, greatly expanded the capabilities of \hst\ for
dynamical measurements.  In addition, the development of techniques to
model the kinematic data in detail led to more robust measurements.
At the center of a disk, the large spatial gradients in rotation
velocity and emission-line surface brightness are smeared out by the
telescope point-spread function (PSF) and by the nonzero size of the
spectroscopic aperture.  Macchetto \etal\ (1997) and van~der~Marel \&
van~den~Bosch (1998) were the first to model the effects of
instrumental blurring on \hst\ gas-kinematic data, and detailed
descriptions of modeling techniques have been given by Barth \etal\
(2001), Maciejewski \& Binney (2001), and Marconi \etal\ (2003).

The feasibility of performing a black hole detection in any given
galaxy can be roughly quantified in terms of \rg, the radius of the
``sphere of influence'' over which the black hole dominates the
gravitational potential of its host galaxy.  This quantity is given by
$\rg = G \mbh / \sigmastar^2$.  Projected onto the sky, and scaled to
typical parameters for an \hst\ measurement, this corresponds to
\begin{equation}
\rg = 0.11 \left( \frac{\mbh}{10^8 \msun} \right)
\left( \frac{200 \mathrm{~km~s^{-1}}}{\sigmastar} \right)^2
\left( \frac{20 \mathrm{~Mpc}}{D} \right)  \mathrm{~~arcsec}.
\end{equation}
Detection of black holes via their influence on the motions of stars
or gas is most readily accomplished when observations are able to
probe spatial scales smaller than \rg, but it should be borne in mind
that this is at best an approximate criterion.  The stellar velocity
dispersion is an aperture-dependent quantity, so there is no uniquely
determined value of \rg\ for a given galaxy.  Even when \rg\ is
unresolved the black hole will still influence the motions of stars
and gas at larger radii.  In stellar-dynamical measurements, it is
possible to obtain information from spatial scales smaller than the
instrumental resolution by measuring higher-order moments of the
central line-of-sight velocity profile, since extended wings on the
velocity profile are the signature of high-velocity stars orbiting
close to the black hole (e.g., van~der~Marel 1994).  With sufficiently
high signal-to-noise ratio, the information contained in the full
line-of-sight velocity profile could be exploited in gas-dynamical
measurements as well, although measurements to date have generally
been performed by fitting models to the first and second moments of
the velocity distribution function (i.e., the mean velocity and
line width at each observed position), or in some cases only to the
mean velocities.

The analysis of a gas-dynamical dataset consists of the following
basic steps.  The galaxy's stellar light profile must be measured,
corrected for dust absorption if necessary, and converted to a
three-dimensional luminosity density.  The stellar mass density is
generally assumed to be axisymmetric or spherically symmetric.  A
model velocity field is computed for the combined potential of the
black hole and the galaxy mass distribution, usually assuming a
spatially constant stellar mass-to-light ratio ($M/L$).  After projecting the
velocity field to a given distance and inclination angle, the model is
synthetically ``observed'' by simulating the passage of light through
the spectrograph optics and measuring the resulting model line
profiles.  Finally, the model fit to the measured emission-line
velocity field is optimized to obtain the best-fitting value of \mbh.
In addition to \mbh, the free parameters in the kinematic model fit
include the disk inclination and major axis position angle, and the
stellar $M/L$; these can be determined from the
kinematic data if observations are obtained at three or more parallel
positions of the spectrograph slit.  Maciejewski \& Binney (2001) have
shown that when the slit is wider than the PSF core, there is an
additional signature of the black hole: at one particular location in
the velocity field, the rotational and instrumental broadening will be
oppositely directed, and they will very nearly cancel, giving a very
narrow line profile.  The location of this feature can be used as an
additional diagnostic of \mbh.  With up-to-date analysis techniques
and high-quality STIS data, it is possible to achieve formal
measurement uncertainties on \mbh\ of order $\sim25\%$ or better for
galaxies with well-behaved disks (e.g., Barth \etal\ 2001).  This
makes the gas-dynamical method very competitive with the precision
that can be achieved by stellar-dynamical measurements.

\begin{figure}
\centering
\scalebox{0.4}{\rotatebox{-90}{\includegraphics{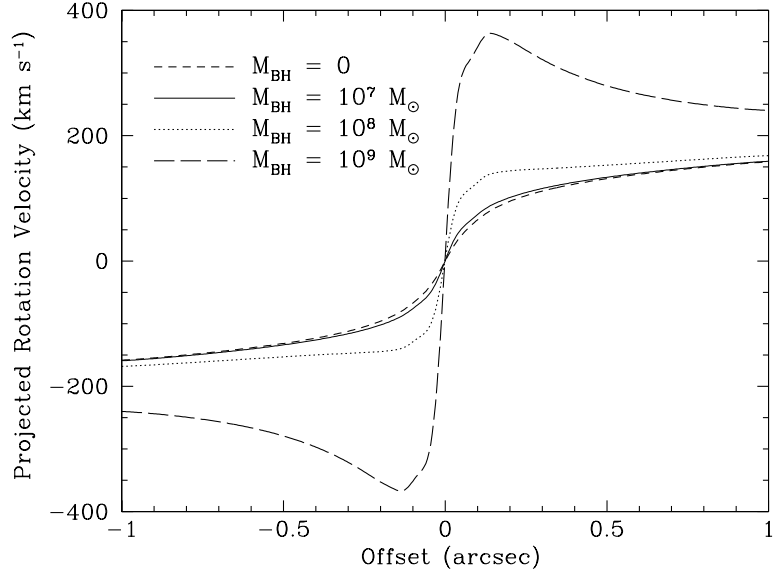}}}
\caption{Projected radial velocity curves for the major axis of a
  model disk with $i = 60\arcdeg$ at $D = 20$ Mpc, for $\mbh = 0$,
  $10^7$, $10^8$, and $10^9$ \msun.  The models are convolved with the
  \hst\ PSF and sampled over a slit width of
  0\farcs1.  The curves illustrate the mean velocity observed at each
  position along the spectrograph slit. }
\label{rotcurves}
\end{figure}

Figure \ref{rotcurves} illustrates model calculations for the
projected radial velocities along the major axis of an inclined gas
disk.  The models have been calculated for a disk inclined at
$60\arcdeg$ to the line of sight, in a galaxy at a distance of 20 Mpc
with $\mbh = 0$, $10^7$, $10^8$, and $10^9$ \msun.  To demonstrate the
effects of varying \mbh, the same stellar mass profile has been used
for all four models.  The model velocity fields have been convolved
with the STIS PSF and sampled over an aperture corresponding to an
0\farcs1-wide slit, and the curves represent the mean velocity that
would be observed as a function of position along the slit.  At one
extreme, the sphere of influence of the $10^7$ \msun\ black hole is
unresolved, and the $10^7$ \msun\ model is barely distinguishable from
the model with no black hole, although the rapidly rotating gas near
the black hole will give rise to high-velocity wings on the central
emission-line profile.  The opposite extreme is the $10^9$ \msun\
black hole, for which the Keplerian region is extremely well resolved
and the black hole would be readily detected.  Instrumental blurring
causes a turnover in the velocity curves at about 0\farcs15 in this
case, and Keplerian rotation can only be clearly detected at larger
radii.  Thus, Keplerian rotation can only be verified in detail (in
the sense of having several independent data points to trace the $\upsilon 
\propto r^{-1/2}$ dependence) for galaxies with exceptionally
well-resolved \rg.  To date, the only published \hst\ gas-dynamical
measurements that have unambiguously detected the Keplerian region are
those of M87 (Macchetto \etal\ 1997) and M84 (Bower \etal\ 1998).

The middle ground between these two extremes is illustrated by the
$10^8$ \msun\ black hole in Figure \ref{rotcurves}.  In this case, the
Keplerian rise in velocity is not detected; instead, a steep but
smooth velocity gradient across the nucleus is observed.  The mass of
the black hole can still be determined from the steepness of this
central gradient, provided that the observations give sufficient
information to distinguish the velocity curves from those of the
$\mbh=0$ case.  It is possible in principle to measure \mbh\ even in
galaxies for which \rg\ is formally unresolved, if it can be shown
that the disk has an excess rotation velocity relative to the
best-fitting model without a black hole.  However, in such cases there
are practical complications that may limit the measurement accuracy:
it is critical to determine the stellar luminosity density profile and
$M/L$ accurately, and spatial gradients in $M/L$ are an
added complication.  In general, the most confident detections of
black holes will be in those relatively rare objects for which the
Keplerian region can be clearly traced, but the majority of
gas-dynamical mass measurements will come from objects for which the
mass is determined primarily from the steepness of the central
velocity gradient rather than from fitting models to a well-resolved
Keplerian velocity field.

Since the gas can respond to nongravitational forces, it is essential
for the observations to map the disk structure in sufficient detail
that the assumption of gravitational motion can be tested.  This is a
nontrivial concern, as there are examples of galaxies in which the gas
does not rotate at the circular velocity (e.g., Fillmore, Boroson, \&
Dressler 1986; Kormendy \& Westpfahl 1989). The case of IC 1459 serves
as a cautionary tale.  FOS observations at 6 positions within 0\farcs3
of the nucleus revealed a steep central velocity gradient in the
ionized gas, and disk model fits implied the presence of a central
dark mass of $(1-4) \times10^8$ \msun\ (Verdoes Kleijn \etal\ 2000).
More recently, STIS observations have mapped out the circumnuclear
kinematics in much greater detail and found an irregular and
asymmetric velocity field that cannot be interpreted in terms of flat
disk models; a stellar-dynamical analysis finds $\mbh =
(2.6\pm1.1)\times10^9$ \msun\ (Cappellari \etal\ 2002).  Thus, the
ionized gas fails to be a useful probe of the gravitational potential
in this galaxy.

Most of the nuclear disks observed spectroscopically with \hst\ have
revealed evidence for substantial internal velocity dispersions, even
when the velocity field is clearly dominated by rotation.  That is,
the gas disks are not in purely quiescent circular rotation, and the
disks have some degree of internal velocity structure or turbulence.
The intrinsic emission-line velocity dispersion \sgas\ tends to be
greatest at the nucleus, and observed values span a wide range,
exceeding 500 \kms\ in extreme cases (e.g., van~der~Marel \& van~den~Bosch 
1998). 

The origin of this internal velocity dispersion is unknown, and the
interpretation of \sgas\ remains the single most important unresolved
problem for gas-dynamical measurements of black hole masses.  In a
study of the radio galaxy NGC 7052, van~der~Marel \& van~den~Bosch
(1998) argued that \sgas\ was due to local turbulence in gas that
remained in bulk motion on circular orbits at the local circular
velocity.  Another possibility that has been considered is that the
disks are composed of a large number of clouds with small filling
factor, and that the individual clouds are on noncircular orbits that
are nevertheless dominated by gravity (e.g., Verdoes Kleijn \etal\
2000).  This is analogous to the effect of ``asymmetric drift'' that
is well known in stellar dynamics (e.g., Binney \& Tremaine 1987).  In
this situation, the gas disk would be supported against gravity by
both rotation and random motions, and models based on pure circular
rotation would underestimate the true black hole masses.  The
asymmetric drift correction to \mbh\ is expected to be small when
$(\sgas / \upsilon_{\mathrm{rot}})^2 \ll 1$, but this must be tested on a
case-by-case basis by computing models for the emission-line widths.
Various approaches to calculating the asymmetric drift in gas disks
have been presented by Cretton, Rix, \& de Zeeuw (2000), Verdoes
Kleijn \etal\ (2000, 2002), and Barth \etal\ (2001).

Since different groups have used a variety of methods to treat this
problem in dynamical analyses (sometimes ignoring it altogether), and
since \sgas\ varies widely among different galaxies, the influence of
the intrinsic velocity dispersion could be responsible for some of the
apparent scatter in the \msigma\ relation.  Now that large samples of
galaxies have been observed with STIS, it may be possible to discern
whether there are any correlations of \sgas\ with the level of nuclear
activity or with the Hubble type or any other property of the host
galaxy; detection of any clear trends might help to elucidate the
origin of the intrinsic dispersion.  Three-dimensional
hydrodynamical simulations have evolved to the point where it is now
becoming feasible to model the turbulent structure of gas disks in
galaxies (e.g., Wada, Meurer, \& Norman 2002); this work may lead to
new insights on how best to interpret \sgas\ in gas-dynamical
measurements.   Further discussion of the intrinsic velocity dispersion
problem is presented by Verdoes Kleijn, van~der~Marel, \& Noel-Storr (2003).

Most gas-dynamical studies to date have concentrated on elliptical and
S0 galaxies, but STIS data have now been obtained for dozens of spiral
galaxies as well (e.g., Marconi \etal\ 2003).  In principle, the
gas-dynamical method can work equally well for spirals, but there are
some additional complications.  One is the presence of nuclear star
clusters, which are nearly ubiquitous in late-type spirals (Carollo,
Stiavelli, \& Mack 1998; B\"{o}ker \etal\ 2002).  The nuclear star
clusters are typically young (Walcher \etal\ 2003), and the dynamical
modeling must take into account the possible radial gradient in $M/L$.
Another caveat is that spiral arms or bar structure can lead to
departures from circular rotation in the ionized gas (Koda \& Wada
2002; Maciejewski 2003).  Only $\sim10\%-20\%$ of spirals appear to have
orderly emission-line velocity fields suitable for disk-model fitting
to derive \mbh\ (Sarzi \etal\ 2001), so some care is needed when
selecting targets for \hst\ gas-dynamical surveys.  Ho \etal\ (2002)
have shown that orderly rotation in the emission-line gas almost
exclusively occurs in galaxies having orderly, symmetric circumnuclear
dust-lane morphology that can be detected in \hst\ imaging data, and
this offers a promising way to maximize the rate of successful \mbh\
measurements in future programs.

Even when gas-dynamical data cannot provide a direct measurement of
black hole mass, either due to insufficient resolution or irregular
kinematics, they can still be used to derive upper limits to \mbh\
from the central amplitude of the rotation curve or the central
emission-line width.  For galaxies with $\sigmastar \lesssim 100$
\kms, where direct measurements of \mbh\ are scarce, upper limits are
valuable additions to the black hole census.  For example, from a
large sample of ground-based rotation curves, Salucci \etal\ (2000)
demonstrated that late-type spirals must generally have $\mbh \lesssim
10^{6-7}$ \msun.  Sarzi \etal\ (2002) measured upper limits to \mbh\
from the central emission-line widths observed in STIS spectra of 16
galaxies having \sigmastar\ between 80 and 270 \kms.  The derived
upper limits are consistent with the \msigma\ relation, further
confirming that few galaxies (perhaps none) are strong outliers with
very overmassive black holes relative to their bulge luminosity or
velocity dispersion.

Gas-dynamical measurements using \hst\ data have now been published
for about a dozen galaxies (Harms \etal\ 1994; Ferrarese \etal\ 1996;
Macchetto \etal\ 1997; Bower \etal\ 1998; van~der~Marel \& van~den~Bosch 1998; 
Ferrarese \& Ford 1999; Barth \etal\ 2001; Sarzi \etal\ 2001; 
Verdoes Kleijn \etal\ 2002; Marconi \etal\ 2003).  The results,
together with stellar-dynamical measurements, have been compiled by
Kormendy \& Gebhardt (2001), Merritt \& Ferrarese (2001), and Tremaine
\etal\ (2002).  This number will continue to increase over the next
several years, since STIS gas-kinematic observations have been
obtained or scheduled for over 100 galaxies.  However, it remains to
be seen how many of these datasets will lead to accurate measurements
of black hole masses.  A substantial fraction of the galaxies will
probably not have well-behaved gas disks suitable for modeling, and in
many of the target galaxies \rg\ may be much smaller than the
resolution limit of \hst\ (Merritt \& Ferrarese 2002).  Thus, it
remains worthwhile to search for additional nearby galaxies having
morphologically regular disks that will make promising targets for
future spectroscopic observations.

With the small projected sizes of \rg\ in most nearby galaxies,
ground-based observations cannot generally be used to perform
gas-dynamical measurements of \mbh.  One exception is the nearby ($D =
3.5$ Mpc) radio galaxy Cen~A.  Marconi \etal\ (2001) obtained
near-infrared spectra of this galaxy in 0\farcs5 seeing, and measured
the velocity field of the Pa$\beta$ and [\ion{Fe}{2}] emission lines.
They detected a steep central velocity gradient with a turnover to
Keplerian rotation outside the region dominated by atmospheric seeing,
and derived a central dark mass of $2^{+3.0}_{-1.4}\times10^8$ \msun.
Although the black hole mass was not determined with very high
precision, this measurement serves as a proof of concept that
ground-based observations of near-infrared emission lines can be used
for gas-dynamical analysis.  With future laser guide star systems,
such measurements can be performed using adaptive optics on 8--10
meter telescopes, although the possibility of a spatially and
temporally variable PSF will make the data analysis a formidable
challenge.  Also, the \emph{Atacama Large Millimeter Array} will
provide new high-resolution views of circumnuclear disks and new
probes of disk dynamics with molecular emission lines.

\section{Black Hole Masses from Observations of H$_2$O Masers}

Water vapor maser emission at $\lambda=1.35$ cm from the nucleus of
the low-luminosity Seyfert 2 galaxy NGC 4258 was detected by Claussen,
Heiligman, \& Lo (1984).  The maser emission consists of a bright core
with several distinct components near the systemic velocity arranged
in an elongated region (Greenhill \etal\ 1995), as well as
``satellite'' lines separated by $\pm800-1000$ \kms\ from the systemic
features (Nakai, Inoue, \& Miyoshi 1993).  The presence of
high-velocity emission was suggestive of a disk rotating about a
central mass of $\sim10^7$ \msun\ (Watson \& Wallin 1994), and the
breakthrough observation came when Miyoshi \etal\ (1995) used the Very
Long Baseline Array to map out the positions of the high-velocity
features.  They found that the satellite masers traced out a
near-perfect Keplerian velocity curve on either side of the nucleus,
allowing a precise determination of the enclosed mass within the inner
radius of the disk (3.9 milliarcsec, corresponding to a radius of only
0.14 pc).  For a distance of 7.2 Mpc, the central mass is found to be
$3.9\times10^7$ \msun.  It is very unlikely that this mass could be
composed of a cluster of dark objects such as stellar remnants or
brown dwarfs, since the cluster density would be so large that its
lifetime against evaporation or collisions would be short compared to
the age of the galaxy (Maoz 1995, 1998).  This makes NGC 4258 the most
compelling dynamical case for the existence of a supermassive black
hole outside of our own Galaxy.  Recently, a preliminary analysis of
\hst\ stellar-dynamical data for NGC 4258 has yielded a value of \mbh\
consistent with the maser measurement (Siopis \etal\ 2002); this is a
reassuring confirmation of the stellar-dynamical technique.

A detailed review of the properties of the NGC 4258 maser system and a
listing of other known \water\ maser galaxies is given by Moran,
Greenhill, \& Herrnstein (1999).  Hundreds of galaxies have been
surveyed for \water\ masers (e.g., Braatz, Wilson, \& Henkel 1996;
Greenhill \etal\ 2002, 2003), and there are about 30 known sources to
date.  Braatz \etal\ (1997) find that powerful \water\ masers are only
detected in Seyfert 2 and LINER nuclei, and not in Seyfert 1 galaxies.
This is consistent with the geometrical picture of AGN unification
models, which posit the existence of an edge-on torus or disk in the
Type 2 objects.  The large path length along our line of sight in an
edge-on disk permits maser amplification, while a more face-on disk in
a Type 1 AGN would not emit maser lines in our direction.  Further
confirmation of this geometric structure comes from detection of large
X-ray obscuring columns (e.g., Makishima \etal\ 1994; Iwasawa,
Maloney, \& Fabian 2002) and polarized emission lines from obscured
nuclei (e.g., Antonucci \& Miller 1985; Wilkes \etal\ 1995) in some
\water\ maser galaxies.

Disklike structures have been detected in several other maser
galaxies, but NGC 4258 remains the only one for which the central mass
has been derived with high precision.  The maser clouds in NGC 1068
appear to trace the surface of a geometrically thick torus rather than
a thin disk, and their velocities fall below a Keplerian curve,
suggesting a central mass of $\sim10^7$ \msun\ (Greenhill \etal\
1996).  NGC 4945 contains high-velocity maser clumps in a roughly
disklike arrangement, but with larger uncertainties than for NGC 4258;
the central mass is $\sim10^6$ \msun\ within a radius of 0.3 pc (Greenhill, 
Moran, \& Herrnstein 1997).  Recently, high-velocity maser features have
been detected in IC 2560 (Ishihara \etal\ 2001), NGC 5793 (Hagiwara
\etal\ 2001), NGC 2960 (Henkel \etal\ 2002), and the Circinus galaxy
(Greenhill \etal\ 2003), and VLBI measurements can determine whether
the masers in these galaxies trace out Keplerian rotation curves.
Additional maser disks will continue to be found in future surveys,
although the nearby AGN population has now been surveyed so thoroughly
that it is unlikely that many more new examples will be found at
distances comparable to that of NGC 4258.

\section{Reverberation Mapping}
\label{reverb}

Reverberation mapping uses temporal variability of active nuclei to
probe the size and structure of the broad-line region (BLR) of Seyfert
1 galaxies and quasars.  The basic principle is that variability in
the ionizing photon output of the central engine will be followed by
corresponding variations in the emission-line luminosity, after a time
delay dependent on the light-travel time between the ionizing source
and the emission-line clouds (Blandford \& McKee 1982).  By monitoring
the continuum and emission-line brightness of a broad-lined AGN over a
sufficient time period, the lag between continuum variations and
emission-line response can be derived, giving a size scale for the
region emitting the line; typical sizes range from a few light-days up
to $\sim1$ light-year.  Thus, the method makes use of the time domain
to resolve structures that cannot be resolved spatially, and the
measurement accuracy depends on temporal sampling rather than spatial
resolution.  Peterson (2001) gives a thorough discussion of the
observational and analysis techniques used in reverberation-mapping
campaigns.  The literature on reverberation mapping is extensive, and
this section reviews only a few recent results on the determination of
black hole masses from reverberation data.

If the motions in the BLR are dominated by gravity (rather than, for
example, radiatively driven outflows), then the central mass can be
derived from the BLR radius combined with a characteristic velocity.
Essentially, the black hole mass is derived as $\mbh = f\upsilon^2r/G$, where
$v$ is some measure of the broad-line velocity width (typically the
full width at half maximum), $r$ is the BLR radius derived from the
measured time delay, and $f$ is an order-unity factor that depends on
the geometry of the BLR (i.e., disklike or spherical).  Reverberation
masses for 34 Seyfert 1 galaxies and low-redshift quasars have been
determined by Wandel, Peterson, \& Malkan (1999) and Kaspi \etal\
(2000).  The method is subject to some potentially serious systematic
errors, however, as emphasized by Krolik (2001); the unknown geometry
and emissivity distribution of the BLR clouds can lead to biases in
the derived masses, and it is critically important to verify that the
BLR velocity field is in fact dominated by gravitational motion.

Recent observations have yielded some encouraging results.  Since the
BLR is radially stratified in ionization level, highly ionized clouds
emitting \ion{He}{2} and \ion{C}{4} respond most quickly to continuum
variations, with response times of a few days, while lines such as
\hbeta\ and \ion{C}{3}] $\lambda1909$ have longer lag times as well as
narrower widths.  Thus, if lag times \tlag\ can be measured for
multiple broad emission lines in a given galaxy, it is possible to
trace out the velocity structure of the BLR as a function of radius.
Peterson \& Wandel (1999, 2000) and Onken \& Peterson (2002) have
shown that for four of the best-observed reverberation targets, NGC
3783, NGC 5548, NGC 7469, and 3C 390.3, the relation between \tlag\
and emission-line width shows exactly the dependence expected for
Keplerian motion.  Within the measurement uncertainties, the emission
lines in each galaxy yield a correlation consistent with FWHM
$\propto$ $\tlag^{0.5}$, and for each galaxy the time lags and
line widths of the different emission lines give consistent results for
the central mass.

Another key question is whether reverberation mapping yields black
hole masses that are consistent with the host galaxy properties of the
AGNs.  Gebhardt \etal\ (2000b) and Ferrarese \etal\ (2001)
demonstrated that the reverberation masses and velocity dispersions of
several Seyfert nuclei are in good agreement with the \msigma\
relation of inactive galaxies.  Nelson (2000) performed a similar
analysis for a larger sample, using [\ion{O}{3}] line widths as a
substitute for \sigmastar, and found similar results, albeit with
additional scatter that could be attributed to some nongravitational
motion in the narrow-line region.  Some previous studies had found the
puzzling result that the AGNs seemed to have systematically lower
\mbh/\mbul\ ratios than inactive galaxies (Ho 1999; Wandel 1999).
Since the AGNs do not appear discrepant in the \msigma\ relation, a
likely conclusion (e.g., Wandel 2002) is that the bulge luminosities
of the Seyferts had been systematically overestimated, due to a
combination of their relatively large distances and the dominance of
the central point sources; starburst activity in the Seyferts might
further bias the photometric decompositions toward anomalously high
bulge luminosities.

Overall, the agreement between the \msigma\ relation of Seyferts with
that of inactive galaxies suggests that the reverberation masses are
probably accurate to a factor of $\sim3$ on average (Peterson 2003).
Direct measurements of \mbh\ in reverberation-mapped Seyferts with
\hst\ would be a valuable cross-check, but unfortunately only a few
bright Seyfert 1 galaxies are near enough for \rg\ to be resolved, and
attempts to perform stellar-dynamical measurements of the nearest
objects with \hst\ have been thwarted by the dominance of the bright
nonstellar continuum.  Gas-dynamical observations are not affected by
this problem, and there have been attempts to detect cleanly rotating
kinematic components in Seyfert narrow-line regions that could be used
for \mbh\ measurements with \hst\ (Winge \etal\ 1999).  Unfortunately,
the presence of outflows or other kinematic disturbances typically
precludes the use of emission-line velocity fields as probes of of
\mbh\ in Seyfert 1 galaxies (e.g., Crenshaw \etal\ 2000).  Since there
is no way to perform direct stellar- or gas-dynamical measurements of
\mbh\ in most reverberation-mapped AGNs, the comparison with the
\msigma\ relation remains the main consistency check that can
currently be applied to the reverberation-based masses.

Reverberation mapping can be extended to higher redshifts, since it is
not dependent on spatial resolution, but the longer variability
time scales for higher-mass black holes in luminous quasars, combined
with cosmological time dilation, can require monitoring campaigns with
durations that are a significant fraction of an individual
astronomer's career!  Kaspi \etal\ (2003) present preliminary results
from an ongoing, eight-year campaign to monitor 11 luminous quasars at
$2.1 < z < 3.2$.  Continuum variations have been detected, but the
corresponding emission-line variability has not yet been seen.
Reverberation observations of quasars are a fundamental
probe of black hole masses at high redshift, and efforts to monitor
additional quasars over a wide redshift range should be encouraged,
despite the long time scales involved.

One important consequence of the reverberation campaigns has been the
detection of a correlation between the BLR radius and the continuum
luminosity, a result that is expected on the basis of simple
photoionization considerations (e.g., Wandel 1997).  With a sample of
34 reverberation-mapped AGNs, Kaspi \etal\ (2000) find $\rblr \propto
L^{0.7}$ using continuum luminosity at 5100 \AA.  This correlation
offers an extremely valuable shortcut to estimate the BLR size, and
black hole mass, in distant quasars.  While reverberation campaigns
require years of intensive observations, $L$(5100 \AA) and
FWHM(\hbeta) can be measured from a single spectrum, and then combined
to yield an estimate of \mbh\ under the assumption of virial motion of
the BLR clouds.  This is the only technique that can routinely be
applied to derive \mbh\ in distant AGNs, and it has recently been the
subject of intense interest, with numerous studies focused on topics
such as the \mbh/\mbul\ ratio in quasars and the search for possible
differences between radio-loud and radio-quiet objects (e.g., Laor
1998, 2001; Lacy \etal\ 2001; McLure \& Dunlop 2001, 2002; Jarvis \& 
McLure 2002; Oshlack, Webster, \& Whiting 2002; Shields \etal\ 2003). 
The technique has also been extended to
make use of continuum luminosity in the rest-frame ultraviolet
combined with the velocity width of either \ion{C}{4} (Vestergaard
2002) or \ion{Mg}{2} (McLure \& Jarvis 2002), so that ground-based,
optical spectra can be used to derive black hole masses for
high-redshift quasars.

These secondary methods are extremely valuable since they offer the
most straightforward estimates of black hole masses at high redshift,
although there are potential biases that must be kept in mind.  The
derived BLR size and \mbh\ depend on the observed continuum
luminosity, but this can be affected by dust extinction (Baker \&
Hunstead 1995) or by relativistic beaming of synchrotron emission in
radio-loud objects (Whiting, Webster, \& Francis 2001).  If the BLR
has a flattened, disklike geometry, then the effects of source
orientation on the observed line width must be accounted for (e.g.,
McLure \& Dunlop 2002).  An additional concern is that the \rblr-$L$
relation has only been calibrated against the Kaspi \etal\
reverberation sample, which covers a somewhat limited range both in
black hole mass ($\mbh \lesssim 5\times10^8$ \msun) and in luminosity
($\lambda L_\lambda$ $\lesssim 7\times10^{45}$ erg s\per\ at 5100
\AA).  Application of this method to high-luminosity quasars with
$\mbh>10^9$ \msun\ necessarily involves a large extrapolation (see
Netzer 2003 for further discussion).  These issues can be overcome if
reverberation masses can be derived for larger samples of quasars,
extending to high intrinsic luminosities and $\mbh > 10^9$ \msun, so
that the relations between \mbh, line width, and luminosity can be
calibrated over a broader parameter space.

\section{Future Work and Some Open Questions}

I conclude with a very incomplete list of a few important and
tractable problems that can be addressed in the foreseeable future by
new observations.

1. More dynamical measurements of black hole masses in nearby galaxies
   are needed, over the widest possible range of host galaxy masses
   and velocity dispersions, so that the slope of the \msigma\ and
   \mbh-\lbul\ correlations can be determined definitively.  To
   constrain the amount of intrinsic scatter in these correlations,
   realistic estimates of the measurement uncertainties are crucial.
   Gas-dynamical measurements with \hst\ and, in the future, from
   ground-based telescopes with adaptive optics, will be a key
   component of this pursuit.  Additional measurements of black hole
   masses from maser dynamics will be extremely valuable as well, if
   more galaxies with Keplerian maser disks can be found.
  
2. Direct comparisons of stellar- and gas-dynamical measurements for
   the same galaxies are a needed consistency check that should be
   performed for galaxies over a wide range of Hubble types and
   velocity dispersions.

3. What causes the intrinsic velocity dispersion observed in nuclear
   gas disks?  Is it possible to determine central masses accurately
   for disks having $(\sigma_\mathrm{gas} / \upsilon_\mathrm{rot})^2 \approx
   1$?  Again, direct comparisons with stellar-dynamical observations
   would be very useful.

4. What can we learn about black hole demographics from AGNs at the
   extremes of the Hubble sequence?  The broad-line widths and
   continuum luminosities of high-redshift quasars imply masses of up
   to $\sim10^{10}$ \msun\ for some objects (e.g., Shields \etal\
   2003), but these measurements involve extrapolating the known
   correlation between \rblr\ and $L$ far beyond the mass and
   luminosity ranges over which it has been calibrated locally.  At
   the other extreme, is there a lower limit to \lbul\ or \sigmastar\
   below which galaxies have no central black hole at all?  Dynamical
   searches for black holes in the nuclei of dwarf ellipticals or very
   late-type spirals become extremely difficult for distances beyond
   the Local Group (see van~der~Marel, this volume).  On the other
   hand, searches for accretion-powered nuclear activity in dwarf
   galaxies can offer some constraints on the population of black
   holes with $M < 10^6$ \msun.  The case of NGC 4395, a dwarf
   Magellanic spiral hosting a full-fledged Seyfert 1 nucleus
   (Filippenko \& Sargent 1989) with a black hole of $\lesssim10^5$
   \msun\ (Iwasawa \etal\ 2000; Filippenko \& Ho 2003), demonstrates
   that at least some dwarf galaxies can host black holes that would
   be undetectable by dynamical means.

5. How do the \msigma\ and $\mbh-\lbul$ correlations evolve with
   redshift, and how early did the black holes in the highest-redshift
   quasars build up most of their mass?  Reverberation mapping of
   high-redshift quasars, and further calibration and testing of the
   \rblr-luminosity relationship in luminous quasars, will be of
   fundamental importance in answering these questions.  Measurement
   of the masses of black holes at high redshift, as well as the
   luminosities and/or velocity dispersions of their host galaxies,
   will be a major observational step toward understanding the
   coevolution of black holes and their host galaxies.

\end{document}